\documentclass[aip,
nofootinbib,
rsi,%
reprint,
twocolumn,
longbibliography,
author-numerical%
]{revtex4-2}

\usepackage{graphicx}
\usepackage{textcomp}

\usepackage{amsfonts}
\usepackage{amssymb}
\usepackage{amsmath}
\usepackage{color}
\usepackage{natbib}

\newcommand{\pdr}[2]{\dfrac{\partial {#1}}{\partial {#2}}}
\newcommand{\pddr}[2]{\dfrac{\partial^2{#1}}{\partial{#2}^2}}

\newcommand{\pdra}[2]{{\partial  {#1}}/{\partial {#2}}}

\newcommand{\tx}{\tilde{x}}
\newcommand{\ty}{y}

\newcommand{\tit}{\tilde{t}}

\newcommand{\hj}{\hat{j}}

\newcommand{\teta}{\tilde{\eta}}

\newcommand{\tJ}{\tilde{J}}

\newcommand{\tX}{\tilde{X}}

\newcommand{\tN}{\tilde{N}}

\newcommand{\tp}{\tilde{p}}
\newcommand{\tR}{\tilde{R}}

\newcommand{\tZ}{\tilde{Z}}

\newcommand{\ph}{\tp_c}

\newcommand{\phref}{p_{h}^{ref}}
\newcommand{\pwref}{p_{w}^{ref}}
\newcommand{\cref}{c_{ref}}

\newcommand{\tom}{\tilde{\omega}}

\newcommand{\Cdl}{C_{dl}}

\newcommand{\lam}{\lambda}

\newcommand{\veps}{\varepsilon}

\newcommand{\sion}{\sigma_i}

\newcommand{\ri}{{\rm i}}

\newcommand{\lexp}[1]{\exp\left(#1\right)}

\newcommand{\ltanh}[1]{\tanh\left(#1\right)}

\newcommand{\etal}{et al.{ }}

\begin{document}

\sf

\title{Impedance of a hydrogen--fed SOFC anode: Analytical and numerical models
       based on the dusty gas transport model}

\author{Andrei Kulikovsky}
\thanks{ECS member}
\email{A.Kulikovsky@fz-juelich.de}

\affiliation{Forschungszentrum J\"ulich GmbH           \\
    Theory and Computation of Energy Materials (IET--3)   \\
    Institute of Energy and Climate Research,              \\
    D--52425 J\"ulich, Germany
}



\date{\today}

\begin{abstract}
Analytical low-current and numerical high--current models for the impedance
of a hydrogen--fed anode of an anode--supported button SOFC are developed.
The models use the dusty gas transport model for the binary H$_2$--H$_2$O
mixture. We show that neglecting the pressure gradient may lead to a severe underestimation
of the effective hydrogen diffusivity in the support layer.
 A least-squares fitting
of the analytical model to a literature spectrum of a button cell is demonstrated.
The analytical impedance allows to indicate traps when using equivalent circuits
with the Warburg finite-length element for fitting experimental spectra.
The model parameters include the Knudsen hydrogen diffusivity, hydraulic permeability,
porosity/tortuosity ratio of the support layer and the ionic conductivity, double layer capacitance,
and HOR Tafel slope of the active layer. All of the above parameters can be obtained
by fitting the models to experimental spectra.
\end{abstract}

\keywords{Dusty Gas Model, SOFC anode, impedance, analytical solution}

\maketitle

\section{Introduction}

Like PEM fuel cells, Solid Oxide Fuel Cells (SOFCs) convert the chemical energy of hydrogen--oxygen
combustion directly into electricity. However, unlike PEMFCs,
SOFCs operate at high temperature, which allows
in--situ conversion of methane to hydrogen and hydrogen to electric power.
Most of the households in developed countries have access to methane, which makes
the development of a stationary 10--20 kW SOFC power generator a very attractive goal.

Electrochemical impedance spectroscopy (EIS) is, perhaps, the most powerful technique
for non--destructive, in-operando analysis and characterization of SOFCs\cite{Lasia_book_14}.
Impedance spectra contain virtually all information on the transport and kinetic
processes inside the cell. However, extracting this information requires quite
sophisticated modeling.

A vast majority of works on SOFC impedance spectroscopy employ equivalent
circuit models (ECMs)\cite{Primdahl_99,Mogensen_03b,Barfod_07,Sonn_08,%
Leonide_09,Kromp_13,Brett_13,Nielsen_14}.
The construction of an electrical circuit with an impedance spectrum close to the experimental one
is a relatively straightforward procedure leading to fast fitting codes.
However, equivalent spectra are not unique and they typically
contain constant phase elements with no clear physical meaning.
Furthermore, the hydrogen transport in the anode support layer (ASL)
is usually described by the Warburg finite-length
impedance. However, the Warburg impedance does not take into account the pressure
gradient effects in the ASL, nor does it account for the finite double layer
capacitance of the active layer, to which the ASL is attached (see below). A strong
criticism of ECM has been made by Macdonald in his seminal paper\cite{Macdonald_06}.

Physics-based impedance models employing classical transient
charge and mass conservation equations are free from these drawbacks.
There are two ways to calculate impedance from these equations.
Applying a small-amplitude harmonic
current or potential perturbation at $t=0$, the transient equations can be solved
in the time domain. The impedance is then calculated from a fast Fourier
transform of the solution \cite{Bessler_05,Zhu_06,Bessler_07c,Shi_08,Hofmann_10}.
Bessler~\cite{Bessler_07} suggested calculating impedance by applying a small
potential step to the cell rather than a harmonic AC signal. The advantage of this
approach is that the entire impedance spectrum can be obtained from a single model
run, as the step-like function contains all the necessary frequencies
in its Fourier spectrum.

Alternatively, the conservation equations can be directly
linearized and Fourier-transformed. In the case of a single spatial dimension,
this leads to a system of linear ODEs for the small perturbation amplitudes, which
can be solved either numerically\cite{Bertei_16,Donazzi_21}, or analytically\cite{Kulikovsky_24g},
if possible. This approach requires some preliminary analytical work
(linearization and Fourier-transform of equations), but results in a much faster numerical code.

Fitting an ECM to an experimental spectrum typically returns the cell resistivities
and the DL capacitance. The great advantage of fitting a physics-based model is that
it also provides the transport coefficients of the porous media.

Zhu and Kee~\cite{Zhu_06} developed one of the most comprehensive numerical impedance
models that takes into account methane reforming in the anode chamber. The transport of gases
in the porous layers has been calculated using the full Dusty--Gas Model (DGM), including
the pressure gradient term. However, the charge--transfer reactions in the active layers
have been described in a simplified manner, using parallel $RC$-circuit elements.
Hofmann and Panopoulos~\cite{Hofmann_10} calculated the cell impedance from
a transient numerical model for SOFC performance based on a commercial CFD solver.
Bessler \etal\cite{Bessler_05} reported the impedance model with the detailed multistep
hydrogen oxidation reaction (HOR) mechanism. In his model, the pressure
gradient effects in the porous layers were neglected.

Shi \etal\cite{Shi_08} developed a one--dimensional through--plane transient model
for SOFC performance and calculated the impedance from a time--domain
solution. The hydrogen transport through
the ASL was described neglecting the pressure gradient.
Later, Shi \etal\cite{Shi_12} extended their model to two spatial dimensions.

Fu \etal\cite{Fu_15} presented analysis of multicomponent diffusion in the porous
anode and compared the impedance spectra measured at OCV with the model calculations.
The authors\cite{Fu_15} showed that the pressure gradient effects can be neglected
if the anode porosity is sufficiently high. However, no criteria for such
a neglect have been reported.

Bertei \etal\cite{Bertei_16} developed a detailed 1d
impedance model with the transport properties of the porous layers determined from
the microstructural model. Their model included the full DGM for the
gaseous transport on both sides of the cell.
Bertei \etal\cite{Bertei_16b} reported an analytical macro-homogeneous impedance
model for the SOFC anode assuming linear kinetics of the faradaic reaction and
neglecting pressure gradient effects. The  finite-length Warburg terms were derived
for hydrogen and water transport in the ASL. This derivation used the Nernst
equation to relate the perturbation amplitudes of the overpotential and gas molar fractions.
However, the Nernst equation is not valid in
non-equilibrium conditions.  A rigorous approach requires the use of the
ionic charge conservation equation, which leads to an $RC$--factor in the Warburg
finite-length formula (section~\ref{sec:anly}).

Recently, Donazzi \etal\cite{Donazzi_21} reported a 1d+1d impedance model of the SOFC
with straight channels on both sides. The transport in the electrodes was
described using a Fick's type relation for the fluxes.
The conservation equations were linearized
and Fourier-transformed, and the resulting linear ODEs were solved numerically.
Knappe and Kulikovsky\cite{Kulikovsky_24g} developed an analytical model for the anode
impedance; however, for simplicity, the pressure-gradient driven transport in the ASL
was ignored.

From this overview it can be seen that only two works\cite{Zhu_06,Bertei_16}
have included the pressure gradient effects in the impedance calculations. Both the
models\cite{Zhu_06,Bertei_16} are numerical and
none of the works clarified the role
of the pressure gradient in hydrogen transport. The effect of the pressure gradient in
the ASL on cell impedance is not fully understood.

In this work, we develop analytical and numerical physics-based models for the anode
impedance of an anode-supported SOFC operating on pure hydrogen. The models employ
DGM equations to describe the H$_2$--H$_2$O mixture transport
in the ASL. The models highlight the role of the pressure gradient
in hydrogen transport through the ASL. The analytical model allows to
discuss the traps when using the Warburg finite--length element in ECMs.

The analytical model is very fast and it can replace the ECMs in fitting
low-current spectra, including those measured at OCV. The slower and more complicated
numerical model could still be used to fit spectra measured at medium to high DC currents.

\section{Transport and charge conservation equations}

The  models are based on the following assumptions.
\begin{itemize}

\item The hydrogen transport loss in the anode active
   layer (AAL) is negligible. This assumption is justified as the AAL
   thickness is typically two orders of magnitude smaller than the support layer thickness.

\item The rate of the HOR is described by the Butler--Volmer
   equation. This is the standard approach in modeling SOFC impedance.

\item  The variation of the electronic phase potential in the anode
    is ignored, since the electron conductivity
    of the nickel cermet is usually several orders of magnitude higher, than the
    ionic conductivity.

\end{itemize}
Additional assumptions specific to the analytical model are
discussed in Section~\ref{sec:anly}.

\begin{figure}
    \begin{center}
        \includegraphics[scale=0.8]{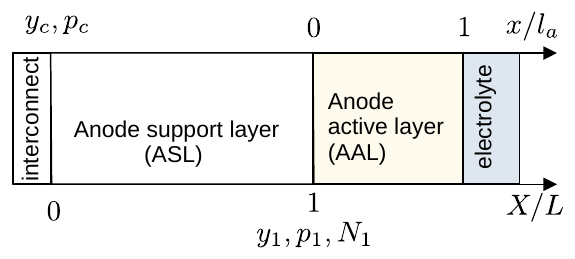}
        \caption{Schematic of the anode--supported SOFC anode. The sketch is strongly
           not to scale: the active layer thickness is two orders of magnitude
           smaller than the ASL thickness.
        }
        \label{fig:scheme}
    \end{center}
\end{figure}

The hydrogen and water transport in the ASL is described
within the scope of the classical DGM:
\begin{multline}
   \sum_{i\neq k}\dfrac{y_i N_k - y_k N_i}{D_{ik}} + \dfrac{N_k}{D_{K,k}} = \\
   {} - \dfrac{p}{RT}\pdr{y_k}{X}
      - \dfrac{1}{RT}\left(y_k + \dfrac{y_k}{D_{K,k}}\dfrac{p B_0 }{\mu}\right)\pdr{p}{X}
      \label{eq:dgm_base1}
\end{multline}
Here
$N_k$ is the molar flux of the $k$th component,
$p$ the pressure,
$D_{i,k}$ the effective binary molecular diffusion coefficient,
$D_{K,k}$ the effective Knudsen diffusion coefficient,
$B_0$ the hydraulic permeability of the porous media,
$\mu$ the mixture kinematic viscosity, and
$X$ is the coordinate through the ASL (Figure~\ref{fig:scheme}).
The DGM takes into account the inter--diffusion Stefan--Maxwell fluxes (the first
term on the left side), the Knudsen diffusion in smaller pores (the second term on the left side), and
the flux due to the pressure gradient (the last term in Eq.\eqref{eq:dgm_base1}).

\subsection{Transport equations in the support layer}

No reactions run in the ASL and hence the transient hydrogen mass conservation
equation is
\begin{equation}
   \dfrac{1}{RT}\pdr{(yp)}{t} + \pdr{N}{X} = 0
   \label{eq:mass}
\end{equation}
where
$y$ is the hydrogen molar fraction,
$p$ the pressure,
$N$ the hydrogen molar flux.
Below, the following dimensionless variables will be used
\begin{multline}
   \tX = \dfrac{X}{L}, \quad \tx = \dfrac{x}{l_a}, \quad \tit = \dfrac{t}{t_*},
    \quad \tp = \dfrac{p}{p_*}, \quad \tN = \dfrac{N}{N_*}, \\
    \hj = \dfrac{j}{j_*}, \quad \tJ = \dfrac{J}{J_*}, \quad \teta = \dfrac{\eta}{b_h}, \quad
    \tom = \omega t_*, \quad \tZ = \dfrac{Z l_a}{\sion}
   \label{eq:dless}
\end{multline}
where
\begin{multline}
   p_* = \dfrac{\mu D_{K,h}}{B_0}, \quad t_* = \dfrac{\Cdl b_h}{i_*},
   \quad N_* = \dfrac{\mu D_{K,h}^2}{R T L B_0}, \\
          j_* = \dfrac{\sion b_h}{l_a}, \quad J_* = 2FN_*,
   \label{eq:char}
\end{multline}
$x$ is the coordinate through the active layer (Figure~\ref{fig:scheme}),
$L$ the ASL thickness,
$l_a$ is the AAL thickness,
$J$ the cell current density,
$\eta$ the positive by convention HOR overpotential,
$b_h$ the HOR Tafel slope,
$D_{K,h}$ the Knudsen hydrogen diffusivity in the ASL,
$\Cdl$ the double layer volumetric capacitance,
$i_*$ the HOR volumetric exchange current density,
$\sion$ the AAL ionic conductivity,
$\omega$ the angular frequency, and
$Z$  the impedance.

With these variables, Eq.\eqref{eq:mass} takes the form
\begin{equation}
   \pdr{(y\tp)}{\tit} + \psi^2\pdr{\tN}{\tX} = 0,
   \label{eq:dmass}
\end{equation}
where
\begin{equation}
   \psi = \sqrt{\dfrac{\Cdl b D_{K,h}}{i_* L^2}}.
   \label{eq:psi}
\end{equation}
The dimensionless DGM equations for $y$ and $\tp$ in the binary
H$_2$--H$_2$O mixture are\cite{Kulikovsky_25a2}:
\begin{align}
   & \pdr{(y\tp)}{\tX} + y\tp\pdr{\tp}{\tX} ={} - (1 + K)\tN
      \label{eq:dgm_hyd} \\
   & \bigl(1 + \tp\,(3 - 2y)\bigr)\pdr{\tp}{\tX} = 2\tN
   \label{eq:dgm_sumd}
\end{align}
where
\begin{equation}
   K = \dfrac{D_{K,h}}{D_m}
   \label{eq:K}
\end{equation}
and $D_m$ is the effective molecular diffusion coefficient of the H$_2$--H$_2$O mixture in the ASL.
Eq.\eqref{eq:dgm_hyd} is the dimensionless DGM equation \eqref{eq:dgm_base1}
for the hydrogen molar fraction/flux.
Eq.\eqref{eq:dgm_sumd} is the sum of Eqs.\eqref{eq:dgm_base1}
for the hydrogen and water $y_w$ molar fractions,
taking into account that $y_w = 1- y$ and the water molar flux $N_w =-N$.

\label{sec:LF}

Application of the small--amplitude harmonic AC current to the cell induces
the small--amplitude response of all transport variables. Mathematically,
equations for this response can be derived by linearization
and Fourier--transform of Eqs.\eqref{eq:dmass}, \eqref{eq:dgm_hyd} and \eqref{eq:dgm_sumd}
using the substitutions
\begin{equation}
   \begin{split}
      & y = y^0(\tX) + y^1(\tX, \tom) \exp(\ri\tom\tit), \quad |y^1| \ll y^0 \\
      & \tp = \tp^0(\tX) + \tp^1(\tX, \tom) \exp(\ri\tom\tit), \quad |\tp^1| \ll \tp^0 \\
      & \tN = \tN^0(\tX) + \tN^1(\tX, \tom) \exp(\ri\tom\tit), \quad |\tN^1| \ll \tN^0
   \end{split}
   \label{eq:Fourier}
\end{equation}
where
the subscripts 0 and 1 mark the static variables and the small perturbation amplitudes,
respectively. Note that $y^1$, $\tp^1$ and $\tN^1$ are the perturbation amplitudes
in the $\tom$--space.
Neglecting terms with the
perturbations product and subtracting the respective static equations we come to
\begin{equation}
   \psi^2\pdr{\tN^1}{\tX} = - \left(\tp^0 y^1 + y^0\tp^1\right)\ri\tom,
   \label{eq:tN1}
\end{equation}
\begin{multline}
   \tp^0\pdr{y^1}{\tX} + \tp^1\pdr{y^0}{\tX}
       + \bigl(y^0\tp^1 + y^1(1 + \tp^0)\bigr)\pdr{\tp^0}{\tX} \\
      + y^0 (1 + \tp^0)\pdr{\tp^1}{\tX} = -(1 + K)\tN^1
      \label{eq:hyd1F}
\end{multline}
\begin{multline}
    \bigl(3\tp^1 - 2(\tp^0 y^1 + \tp^1 y^0)\bigr)\pdr{\tp^0}{\tX} \\
     + \bigl(1 + \tp^0 (3 - 2 y^0)\bigr)\pdr{\tp^1}{\tX} = 2\tN^1
      \label{eq:sumd1F}
\end{multline}
Multiplying Eq.\eqref{eq:hyd1F} by 2 and summing with Eq.\eqref{eq:sumd1F}, we get
a more compact equation for $\pdra{y^1}{\tX}$:
\begin{multline}
  2\tp^0\pdr{y^1}{\tX} + 2\tp^1\pdr{y^0}{\tX} + \left(2 y^1 + 3\tp^1\right)\pdr{\tp^0}{\tX} \\
  + \left(1 + 2y^0 + 3\tp^0\right) \pdr{\tp^1}{\tX} = - 2 K \tN^1
  \label{eq:hyd1F2}
\end{multline}

For the static shapes $y^0$ and $\tp^0$ we use the approximate analytical
solutions to the system \eqref{eq:dgm_hyd}, \eqref{eq:dgm_sumd}  \cite{Kulikovsky_25a2}:
\begin{multline}
   y^0 = \dfrac{1}{\ph^0 + W\tX}\biggl(y_c\ph^0 - \dfrac{3}{4} W^2\tX^2 \\
   - \left(\dfrac{(3\ph^0 + 1)}{2} W + K\tJ\right)\tX\biggr)
   \label{eq:y0x}
\end{multline}
\begin{equation}
    \tp^0 = \ph^0 + W\tX
    \label{eq:tp0x}
\end{equation}
where $y_c$, $\ph$ are the parameters in the anode chamber, and
\begin{equation}
   W =  \dfrac{2 \tJ}{1 + \ph (3 - 2 y_c)}.
   \label{eq:W}
\end{equation}
The derivative $\pdra{y^0}{\tX}$ is given by Eq.\eqref{eq:dydx} and $\pdra{\tp^0}{\tX} = W$.

The boundary conditions for Eqs.\eqref{eq:tN1}, \eqref{eq:sumd1F}, \eqref{eq:hyd1F2} are
\begin{equation}
   \tp^1(0) = y^1(0) = 0, \quad \tN^1(0) = \tJ^1
   \label{eq:bcs}
\end{equation}
The first two conditions express
unperturbed pressure and hydrogen molar fraction in the anode channel and the last one means
that $\tN^1(0)$ is equal to the applied current density perturbation $\tJ^1$.
The  system of Eqs.\eqref{eq:tN1}, \eqref{eq:sumd1F}, \eqref{eq:hyd1F2}
with the boundary conditions \eqref{eq:bcs} form a linear initial--value (Cauchy) problem for
$\tN^1$, $\ty^1$ and $\tp^1$ in the ASL.

\subsection{Ionic charge conservation in the active layer}

Due to the small AAL thickness, we neglect the variation of hydrogen and water
concentrations through the AAL depth. The ionic charge conservation equation in the AAL is
\begin{multline}
   \Cdl\pdr{\eta}{t} - \sion\pddr{\eta}{x} \\
       = - i_*\left(\dfrac{p_h}{\phref}\lexp{ \dfrac{\eta}{b_h}}
              - \dfrac{p_w}{\pwref}\lexp{- \dfrac{\eta}{b_w}}\right)
   \label{eq:etax}
\end{multline}
where
$\Cdl$ is the volumetric double layer capacitance,
$\sion$ the AAL ionic conductivity,
$b_w$ the Tafel slope of the hydrogen evolution reaction (HER).
The right side is the Butler--Volmer reaction rate, where
the first exponent describes the direct HOR rate and the second exponent represents
the rate of the reverse HER.

With the dimensionless variables Eqs.\eqref{eq:dless},
Eq.\eqref{eq:etax} takes the form
\begin{multline}
   \pdr{\teta}{\tit} - \veps^2\pddr{\teta}{\tx} = - q\tp\,\bigl(y\exp(\teta) \\
       - r_{ref} (1-y)\exp(- \beta\teta)\bigr)  = - q Q.
   \label{eq:tetax2}
\end{multline}
Here $\beta$, $\veps$, $q$ and $r_{ref}$ are the constant parameters:
\begin{equation}
   \beta = \dfrac{b_h}{b_w}, \quad \veps = \sqrt{\dfrac{\sion b_h}{i_* l_a^2}}
   \label{eq:betaveps}
\end{equation}
\begin{equation}
   q = \dfrac{\mu D_{K,h}}{B_0 p_{h, ref}}, \quad
    r_{ref} = \dfrac{p_{h,ref}}{p_{w,ref}}.
   \label{eq:qrref}
\end{equation}

Linearization of the factor $Q$ in Eq.\eqref{eq:tetax2} gives
\begin{equation}
   Q^1 =  Y y^1 + P \tp^1 + E \teta^1
   \label{eq:tR}
\end{equation}
where
\begin{equation}
   \begin{split}
       & Y = \tp^0\left(\exp(\teta^0) + r_{ref}\exp(-\beta\teta^0)\right) \\
       & P =  y^0\exp(\teta^0) - r_{ref}\exp(-\beta\teta^0) (1 - y^0) \\
       & E = \tp^0\left(y^0\exp(\teta^0) + r_{ref}\beta\exp(- \beta\teta^0) (1 - y^0)\right)
   \end{split}
   \label{eq:YPE}
\end{equation}
Substituting into Eq.\eqref{eq:tetax2} the Fourier--transform
\begin{equation}
   \teta = \teta^0(\tx) + \teta^1(\tx, \tom) \exp(\ri\tom\tit), \quad |\teta^1| \ll \teta^0
\end{equation}
expanding exponents, neglecting the terms with perturbation products and
subtracting the static equation for $\teta^0$, we get the problem for
the overpotential perturbation amplitude $\teta^1$, Eq.\eqref{eq:teta1x}
\begin{multline}
 \veps^2\pddr{\teta^1}{\tx}  = \left(\ri\tom + q E_1\right)\teta^1
     + q \left(Y_1 y_1^1 + P_1\tp_1^1\right), \\
  \left.\pdr{\teta^1}{\tx}\right|_{\tx=0} = 0, \quad \left.\pdr{\teta^1}{\tx}\right|_{\tx=1} = \hj_1^1
 \label{eq:teta1x}
\end{multline}

The terms $Y_1 y_1^1$ and $P_1\tp_1^1$ in Eq.\eqref{eq:teta1x}
include solutions $y_1^1$ and $\tp_1^1$
of the problem \eqref{eq:tN1}, \eqref{eq:sumd1F}, \eqref{eq:hyd1F2}
calculated at the ASL/AAL interface (at $\tX=1$). The
factors $Y_1$, $P_1$ and $E_1$ are given by Eqs.\eqref{eq:YPE} with the
static parameters $y_1^0$, $\tp_1^0$ calculated at $\tX=1$.

The left boundary condition in Eq.\eqref{eq:teta1x}
means zero ionic current through the ASL/AAL interface ($\tx=0$), and
$\hj^1$ is the ionic current density equivalent to the stoichiometric hydrogen flux entering
the AAL/ASL interface
\begin{equation}
   \hj^1 = \dfrac{J_*\tN_1^1}{j_*}
   \label{eq:hj1}
\end{equation}
Note that $\hj^1$ is scaled to satisfy the Ohm's law
in the form of $\hj = \pdra{\teta}{\tx}$ used in Eq.\eqref{eq:tetax2}
and in the boundary condition to Eq.\eqref{eq:teta1x}.

\section{Impedance models}

\subsection{Numerical impedance model}


A high--current numerical anode impedance $Z_{num}$ is calculated
taking into account the variation of the pressure and hydrogen concentration through the ASL
and the variation of the static overpotential $\teta^0$ along $\tx$. The
procedure is as follows.
\begin{enumerate}

\item Fix the static cell current density $\tJ^0$  and solve the system
   \eqref{eq:tN1}, \eqref{eq:sumd1F}, \eqref{eq:hyd1F2} with
   the boundary conditions Eq.\eqref{eq:bcs} and
   with the static shapes $y^0(\tX)$ and $\tp^0(\tX)$
   given by Eqs.\eqref{eq:y0x}, \eqref{eq:tp0x}.

\item
   Solve the static problem for $\teta^0(\tx)$:
      \begin{multline}
         \hphantom{eeeee} \veps^2\pddr{\teta^0}{\tx} = q\tp_1^0\,\bigl(y_1^0\exp(\teta^0) \\
             - r_{ref} (1 - y_1^0)\exp(- \beta\teta^0)\bigr), \\
             \left.\pdr{\teta^0}{\tx}\right|_{\tx=0} = 0,
             \quad \left.\pdr{\teta^0}{\tx}\right|_{\tx=1} = \tJ_0 \dfrac{J_*}{j_*}
           \label{eq:vccnum}
      \end{multline}

\item Solve the problem for $\teta^1$, Eq.\eqref{eq:teta1x}, with $Y$, $P$ and $E$ from Eqs.\eqref{eq:YPE}
   and $\teta^0(\tx)$ from the previous step.

\item Calculate the numerical impedance
    \begin{equation}
       \tZ_{num} = \left.\dfrac{\teta^1}{\pdra{\teta^1}{\tx}}\right|_{\tx=1}.
       \label{eq:tZnum}
    \end{equation}

\end{enumerate}

\subsection{Analytical low-current impedance model}
\label{sec:anly}

When the DC current is small, we can neglect the variation of the static
pressure $\tp^0$ and hydrogen concentration $y^0$ through the ASL depth, but keep
the perturbation amplitudes $y^1$, $\tp^1$, and $\tN^1$  as functions
of $\tX$. The static overpotential $\teta^0$ can also
be approximated by a constant value.
Under these assumptions, the analytical anode impedance $\tZ_{anly}$ can be derived (Appendix~\ref{sec:deriv})
\begin{equation}
   \tZ_{anly} = \dfrac{1}{\phi\tanh\phi}
      + \dfrac{\chi \tR_W \ltanh{\sqrt{\ri\tom\xi/\psi^2}}}{\left(\ri\tom + q E_1\right)\sqrt{\ri\tom\xi/\psi^2}}
   \label{eq:tZanly}
\end{equation}
where
\begin{equation}
   \phi = \dfrac{\sqrt{\ri \tom + q E_1}}{\veps}, \quad
      \chi = \dfrac{q j_*}{J_*}, 
   \label{eq:phi}
\end{equation}
\begin{equation}
   \xi = \dfrac{((3 - 2y_c) K + 3)\ph + K + 1}{1 + \ph (3 - 2y_c)},
   \label{eq:xi}
\end{equation}
 $\psi$ is given in Eq.\eqref{eq:psi}, and
\begin{equation}
   \tR_W = \dfrac{Y_1\xi}{\ph}
       + \dfrac{2 (Y_1 y_c - P_1\ph)}{(1 + \ph (3 - 2y_c))\ph}.
   \label{eq:tRW}
\end{equation}
(see also Table~\ref{tab:dless} for the dimensionless parameters in this work).

The first term in Eq.\eqref{eq:tZanly} is the combined faradaic and ionic transport
impedance of the AAL. The second therm in Eq.\eqref{eq:tZanly} is the ASL transport impedance
\begin{equation}
   \tZ_{tr} = \dfrac{\chi \tR_W \ltanh{\sqrt{\ri\tom\xi/\psi^2}}}
                    {\left(\ri\tom + q E_1\right)\sqrt{\ri\tom\xi/\psi^2}},
   \label{eq:tZtra}
\end{equation}
given by the product of the Warburg finite-length impedance
\begin{equation}
   \tZ_{W} = \dfrac{\chi \tR_W \ltanh{\sqrt{\ri\tom\xi/\psi^2}}}{\sqrt{\ri\tom\xi/\psi^2}}
      \label{eq:tZW}
\end{equation}
and the parallel $RC$-circuit impedance
\begin{equation}
   \tZ_{RC} = \dfrac{1}{\ri\tom + qE_1}
   \label{eq:tZrc}
\end{equation}
The $RC$-factor
arises since the transport layer is attached to the porous active layer
with the finite double layer capacitance. This capacitance (and the
displacement current) were ignored
by Warburg in his classic derivation of the semi--infinite transport layer
impedance\cite{Warburg_1899} and later
in the derivation of the Warburg finite-length impedance.

It can be shown that in the
limit of $\Cdl \to 0$, Eq.\eqref{eq:tZtra} reduces to the Warburg
impedance. Indeed, in Eq.\eqref{eq:tZtra}, the ratio $\tom/\psi^2$,
and the factors $\chi$, $q$, $E_1$, $\tR_W$
are all independent of $\Cdl$. The only term proportional to $\Cdl$ is $\ri\tom$ in the
denominator of Eq.\eqref{eq:tZtra}. Thus,  at $\Cdl \to 0$
the $RC$--impedance reduces to the constant real value $1/(qE_1)$ and Eq.\eqref{eq:tZtra}
transforms to the scaled Warburg impedance $\tZ_{W}/(qE_1)$
(see\cite{Kulikovsky_17j} for further discussions).

In the limit of $\omega \to 0$, the factor $\ltanh{\sqrt{\ri\tom\xi/\psi^2}}/\sqrt{\ri\tom\xi/\psi^2} \to 1$
and from Eq.\eqref{eq:tZtra}
we get the anode transport resistivity $\tR_{tr}$ at low cell current density
\begin{equation}
   \tR_{tr} = \dfrac{\chi\tR_W}{q E_1} = \dfrac{j_*\tR_W}{J_* E_1}
   \label{eq:tRtra}
\end{equation}
Dimension formulas for $R_{tr}$ and $Z_{anly}$ can be obtained
from Eqs.\eqref{eq:tRtra} and \eqref{eq:tZanly},
respectively, using the list of the dimensionless parameters and variables in Table~\ref{tab:dless}.

\section{Numerical results and discussion}

\subsection{Model spectra}

A custom Python code for numerical calculations has been used.
Eqs.\eqref{eq:tN1}, \eqref{eq:sumd1F}, \eqref{eq:hyd1F2} with the boundary conditions
\eqref{eq:bcs} form a complex-valued Cauchy problem, which has been solved by
the standard Runge--Kutta {\em solve\_ivp} solver.
The current--voltage relation, Eq.\eqref{eq:vcc} has been solved using the
{\em fsolve} procedure. Eq.\eqref{eq:vccnum} for the static overpotential, and
the system of real and imaginary parts of Eq.\eqref{eq:teta1x}
for the overpotential perturbation amplitude are the boundary--value problems,
which have been solved using the BVP {\em solve\_bvp} solver.
Unfortunately, {\em solve\_bvp} fails to directly solve the complex--valued BVP
Eq.\eqref{eq:teta1x} for sufficiently high frequencies.

\begin{figure}
\begin{center}
    \includegraphics[scale=0.45]{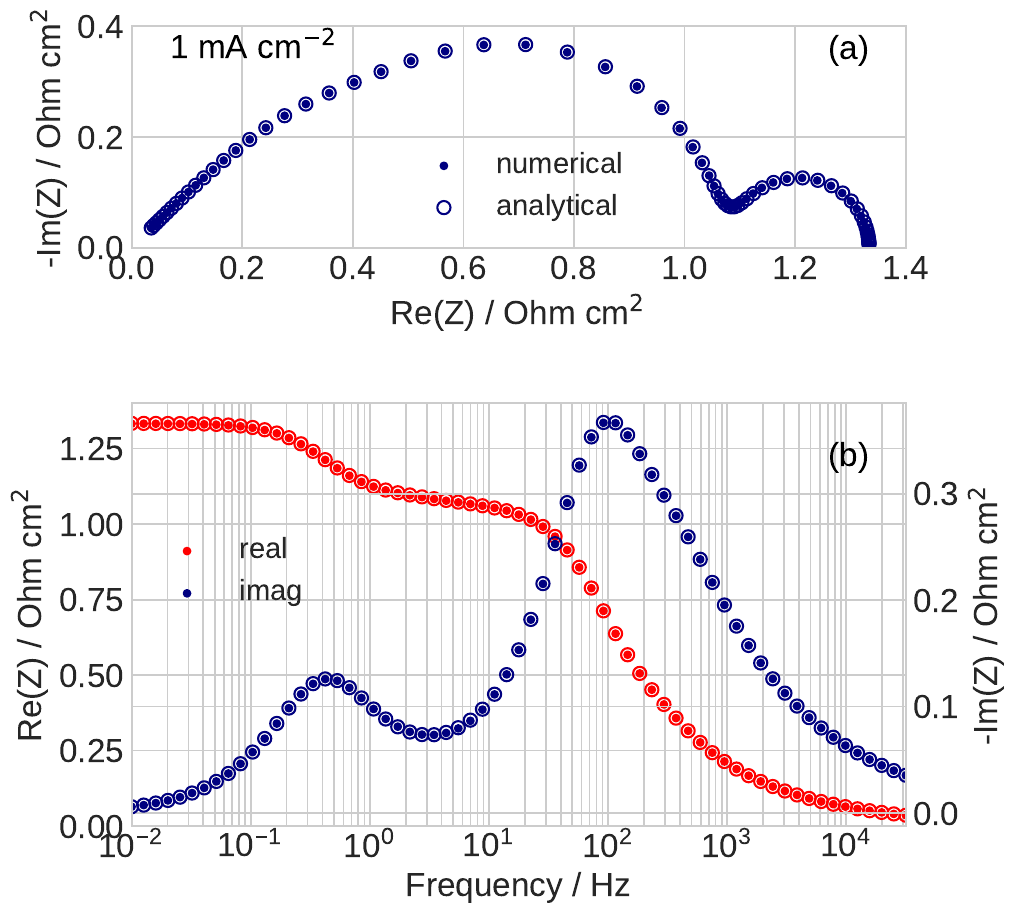}
    \caption{(a) The Nyquist spectra of the numerical anode impedance, Eq.\eqref{eq:tZnum},
    (solid points), and the analytical impedance, Eq.\eqref{eq:tZanly},
     (open circles) for the cell current density of 1 mA~cm$^{-2}$. The
     other parameters are listed in Table~\ref{tab:parms}.
    (b) The Bode plots of the real and imaginary parts of impedance in (a).
    }
    \label{fig:fanly}
\end{center}
\end{figure}

The transport coefficients have been calculated from Eqs.\eqref{eq:tcoeffs}.
The cell parameters taken from the literature are listed in Table~\ref{tab:parms}.
Comparison of the analytical low-current impedance, Eq.\eqref{eq:tZanly},
with the numerical one is shown in Figure~\ref{fig:fanly}. Both the impedances
are calculated at the small current density of 1 mA~cm$^{-2}$, which mimics the
electronic leakage current density in the cell at OCV.
As can be seen, the analytical and the exact numerical spectra are indistinguishable.

\begin{figure}
\begin{center}
    \includegraphics[scale=0.45]{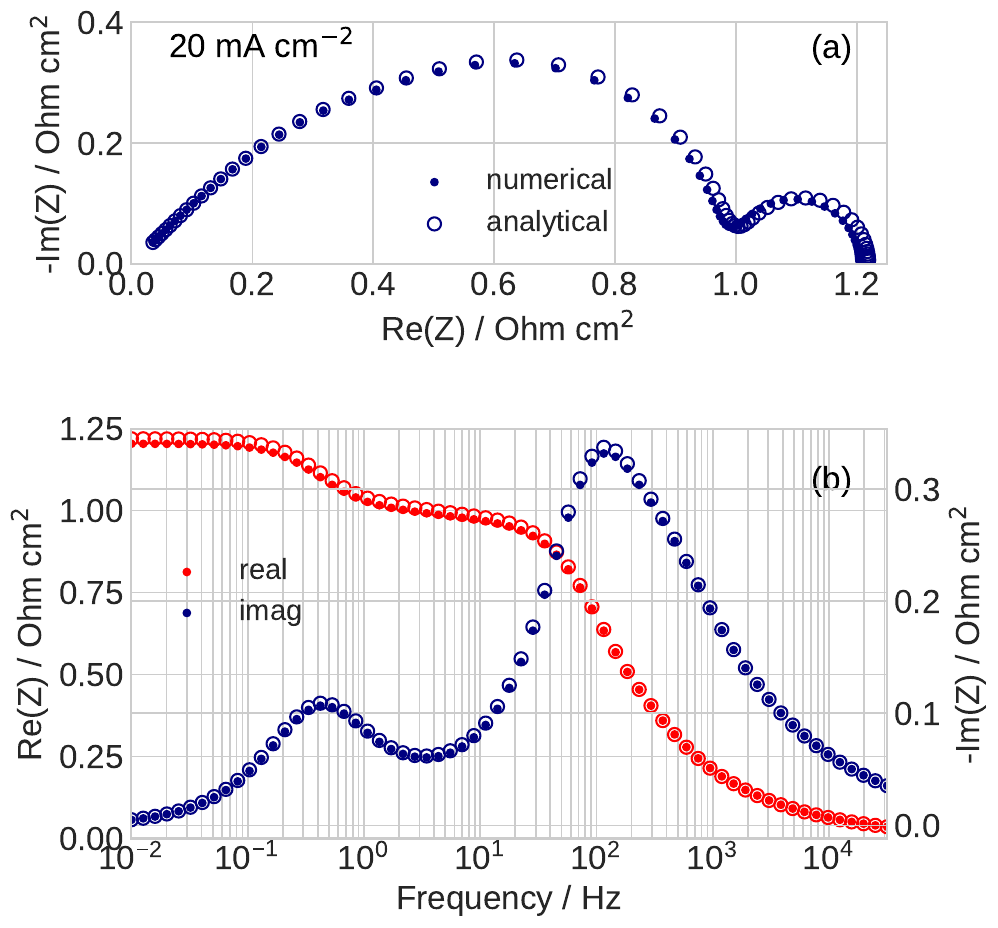}
    \caption{(a) The Nyquist spectra of the numerical anode impedance, Eq.\eqref{eq:tZnum},
    (solid points), and the analytical impedance, Eq.\eqref{eq:tZanly},
    (open circles) for the current density of 20 mA~cm$^{-2}$ and
    the cell parameters in Table~\ref{tab:parms}.
    (b) The Bode plots of the real and imaginary parts of impedance in (a).
    }
    \label{fig:spec1}
\end{center}
\end{figure}

Figure~\ref{fig:spec1} shows the numerical and analytical model spectra
for the cell current density of 20 mA~cm$^{-2}$ and the set of parameters
in Table~\ref{tab:parms}. The analytical model quite well
describes the exact numerical spectrum up to 20~mA~cm$^{-2}$.
Eqs.\eqref{eq:tZanly} can thus be used for fast fitting the
impedance spectra measured at low currents, down to OCV conditions.
Eq.\eqref{eq:tRtra} can be used for calculation of the anode transport
resistivity $R_{tr}$.

The numerical model allows to rationalize the effect of pressure gradient on the anode impedance.
For this purpose, the terms with $\tp^1$, $\pdra{\tp^1}{\tx}$ and $\pdra{\tp^0}{\tx}$
have been set to zero. The reduced numerical model describes the isobaric
anode impedance with purely diffusive hydrogen transport through the ASL.

\begin{figure}
\begin{center}
    \includegraphics[scale=0.45]{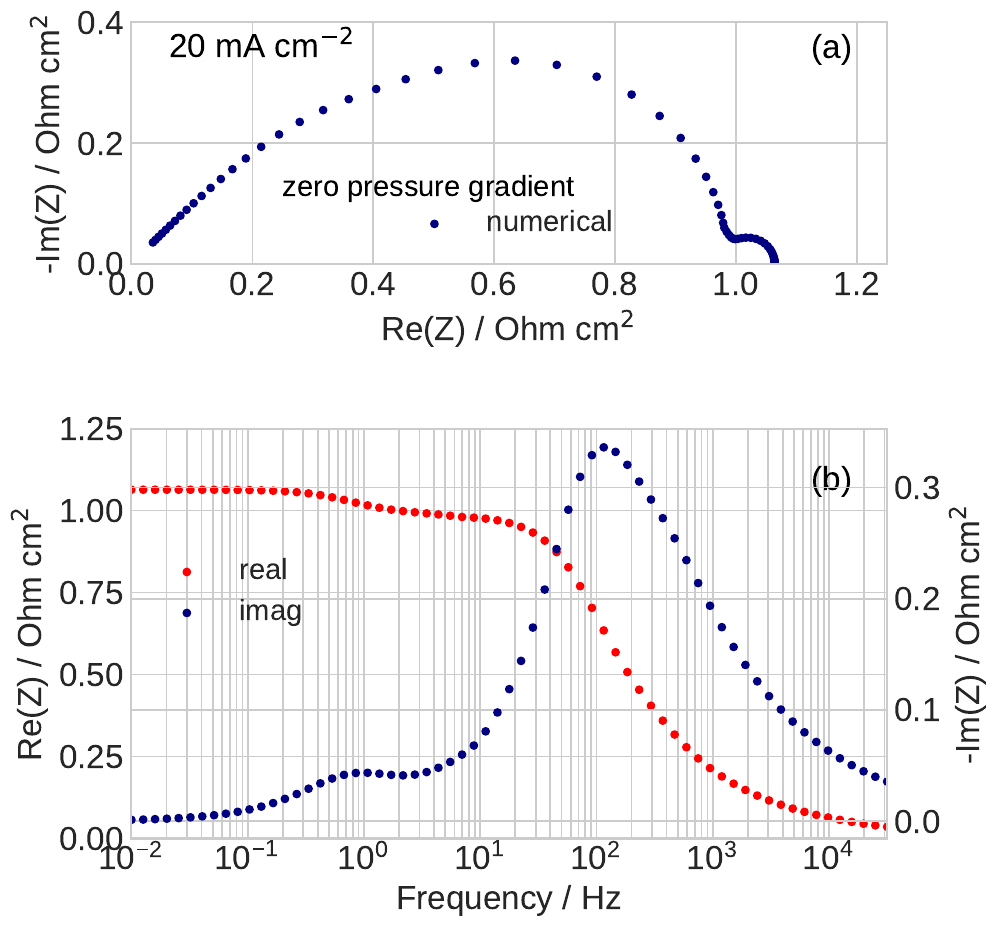}
    \caption{(a) The Nyquist spectrum of the numerical anode impedance, Eq.\eqref{eq:tZnum},
    corresponding to the zero pressure gradient in the anode.
    The current density is 20 mA~cm$^{-2}$ and the cell parameters are listed in in Table~\ref{tab:parms}.
    (b) The Bode plots of the real and imaginary parts of impedance in (a).
    }
    \label{fig:zero_dpdx}
\end{center}
\end{figure}

The resulting spectra
are shown in Figure~\ref{fig:zero_dpdx}. A comparison of Figures~\ref{fig:spec1} and \ref{fig:zero_dpdx}
shows that in spite of the low current density,
the effect of the finite pressure gradient on the spectrum is quite significant:
at zero $\pdra{p}{x}$, the transport arc becomes more than twice smaller
(cf. Figures~\ref{fig:spec1}a and \ref{fig:zero_dpdx}a).
Note that the peak of the transport arc imaginary part
in Figure~\ref{fig:zero_dpdx}b is shifted to twice higher frequency as compared to
Figure~\ref{fig:spec1}b. Both the changes could be ``corrected''
in the isobaric model by using twice lower effective hydrogen diffusivity
in the ASL. In other words, fitting the isobaric impedance model to the experimental
spectra could give a greatly underestimated effective hydrogen diffusivity.

\begin{table*}
\small
\begin{tabular}{|l|c|c|}
\hline
    Cell temperature, K                              & $T$       &  273 + 800   \\
    Pressure in the anode chamber, Pa                & $p_c$     &  $10^5$      \\
    Current density, A~m$^{-2}$                      & $J$       &  $(20\cdot 10^{-3})\cdot 10^4$      \\
    ASL thickness, m                                 & $L$       &  $10^3\cdot 10^{-6}$   \\
    AAL thickness, m                                 & $l_a$     &  $15\cdot 10^{-6}$   \\
    Mean pore diameter, m                            & $d$       &  $0.96\cdot 10^{-6}$ Ref.\cite{Shi_07a}  \\
    Porosity/tortuosity ratio                        & $\lambda$ & 0.033, Ref.\cite{Bao_07} \\
    AAL ionic conductivity@800$^\circ$C, S~m$^{-1}$  & $\sion$   & 0.1 \\
    Double layer capacitance, F~m$^{-3}$             & $\Cdl$    &  $2\cdot 10^6$  \\
    HOR exchange current density A~m$^{-3}$          & $i_*$     &  $10^8$  \\
    HOR Tafel slope, V/exp                           & $b_h$     &   0.1    \\
    Parameter $\beta = b_h/b_w$                      & $\beta$   &  1/3, Ref.\cite{Zhu_06b}  \\
    Reference H$_2$ pressure,                        & $\phref$  & $0.125 p_c$ \\
    Reference water pressure                         & $\pwref$  & $0.042 p_c$    \\
    Hydrogen viscosity@800$^\circ$C, Pa~s    & $\mu$ & $2\cdot 10^{-5}$ \\
    H$_2$/H$_2$O molecular diffusivity m$^2$~s$^{-1}$& $D_m^{free}$ &  $8.154 \cdot 10^{-4}$, Ref.\cite{Bao_07}  \\
    Anode gas composition                            &           &  85\%H$_2$ + 15\%H$_2$O \\
\hline
\end{tabular}
\caption{The cell parameters used in the calculations.
   }
\label{tab:parms}
\end{table*}

\begin{figure}
\begin{center}
    \includegraphics[scale=0.45]{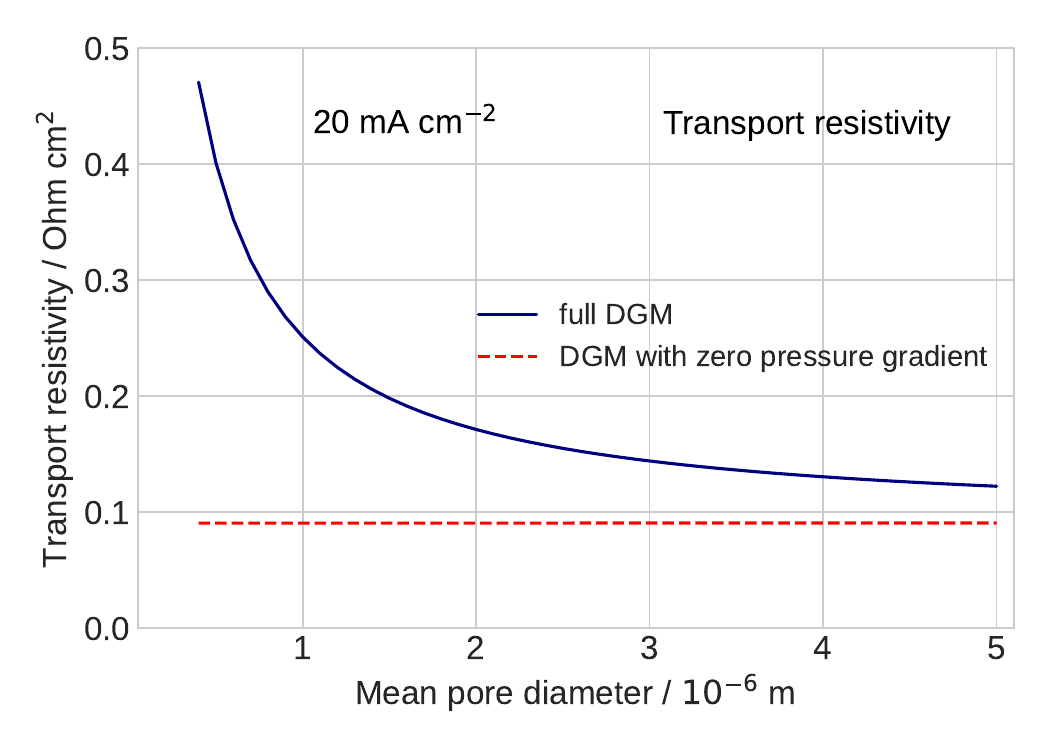}
    \caption{The dependence of transport resistivity on the mean pore diameter in the ASL.
    Solid line -- the full DGM, dashed line -- the reduced DGM with
    the zero pressure  gradient.
    }
    \label{fig:Rtra}
\end{center}
\end{figure}

Figure~\ref{fig:Rtra} shows the ASL transport resistivity $R_{tr}$ calculated from the full DGM
and from the reduced DGM with zero pressure gradient vs the mean pore diameter $d$ in the ASL.
As can be seen, the full DGM
shows the strong effect of $d$ on $R_{tr}$, while the approximation of
zero pressure gradient makes $R_{tr}$ practically insensitive to $d$.
Again we note that any point on the full DGM curve in Figure~\ref{fig:Rtra} could
be obtained within the scope of the zero pressure gradient approximation by
lowering the ASL hydrogen diffusivities, i.e., the parameters $D_{K,h}$ and/or $D_m$
derived from the models neglecting $\pdra{p}{X}$ could be underestimated.

\subsection{Fitting the experimental spectrum}

To demonstrate the analytical model at work, it
has been fitted to the impedance spectrum of the anode--supported button
cell measured by Shi \etal\cite{Shi_08}. The majority of experimental papers present
Nyquist spectra; however, they do not include the complete impedance data necessary
for fitting. Ideally, the frequency dependence of the real and imaginary parts should
be demonstrated. The work of Shi \etal\cite{Shi_08} is one of the very few papers, where the necessary
data are reported.

The spectrum in\cite{Shi_08} was acquired
at open--circuit and the operating conditions collected in Table~\ref{tab:Shi_fixed}.
To describe the high--frequency features of the spectrum, the following model
impedance $Z_a$ has been fitted to the experiment:
\begin{equation}
   Z_a = \dfrac{l_a}{\sion}\tZ_{anly} + \ri \omega L_{cab} S_{cell} + R_{HFR}
   \label{eq:Za}
\end{equation}
where
$\tZ_{anly}$ is given by Eq.\eqref{eq:tZanly},
$L_{cab}$ the cable inductance,
$S_{cell}$ the cell active area, and
$R_{HFR}$ the high--frequency cell resistivity.
The cathode impedance has been neglected\cite{Bertei_16}.
The fitting has been performed using the Python constrained {\em least\_squares} procedure.

\begin{figure}
    \begin{center}
        \includegraphics[scale=0.45]{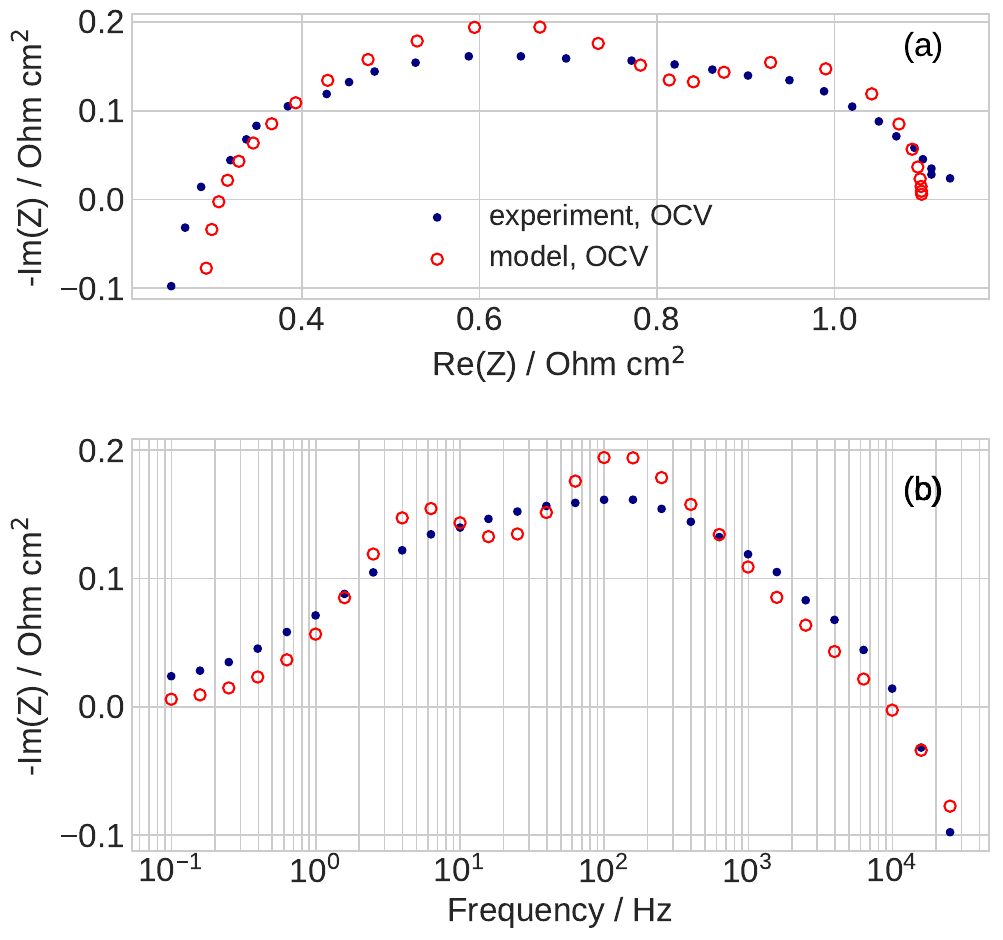} \\
        \caption{ (a) The experimental (solid points)
        and fitted analytical model, Eq.\eqref{eq:tZanly}, (open circles)
         Nyquist spectra of a button
        SOFC at open--circuit potential.
        The experimental points are digitized from Figure 6 of Shi \etal\cite{Shi_08}.
        The cell operating parameters are listed in Table~\ref{tab:Shi_fixed}.
        (b) The Bode plot of imaginary part of impedance in (a).
        }
        \label{fig:Shi}
    \end{center}
\end{figure}

The quality of the spectrum fitting is not high; however, the model correctly captures
the faradaic and transport peaks (Figure~\ref{fig:Shi}).
The fitting parameters are listed in Table~\ref{tab:Shi}.
All the anode parameters are close to their literature values (see discussion in\cite{Kulikovsky_24g}).
The present model returns
the estimate for the porosity/tortuosity ratio $\lam \simeq 0.076$ (Table~\ref{tab:Shi}),
which is nearly twice the value of $0.043$ following from the data reported
by Shi \etal\cite{Shi_08}. It should be noted that the parameters $b_h$,
$r_{ref}$ and porosity/tortuosity ratio strongly depend
on the HOR exchange current density $i_*$. The latter has been fixed here using
Eq.\eqref{eq:iast}\cite{Leonide_09}:
\begin{equation}
   i_* = \dfrac{\gamma}{l_a} y_c^a (1-y_c)^b \lexp{-\dfrac{E_{act}}{RT}}, \quad \text{A m$^{-3}$}
      \label{eq:iast}
\end{equation}
where $\gamma=1.83\cdot 10^6 T$, $a=-0.1$, $b=0.33$, $E_{act}=105.04$ kJ~mol$^{-1}$.

The results presented in Table~\ref{tab:Shi} should be considered as estimates.
More accurate and reliable data could be obtained through the analysis of several spectra.
The spectrum in Figure~\ref{fig:Shi} could, of course, be fitted using the numerical model.
In this case, the Python code runtime is about 33 s on a standard notebook, which
is three orders of magnitude larger than the fitting with the analytical model.

\begin{table}
    \small
\begin{center}
\begin{tabular}{|l|c|}
    \hline
    Cell temperature $T$, K                      & 273 + 800       \\
    Total anode pressure $p$, kPa                & 101.325         \\
    Hydrogen partial pressure $p_h$,             & $0.958 p$       \\
    Reference H$_2$ pressure, $p_{h,ref}$        & $0.125 p$       \\
    Water vapor partial pressure $p_{w}$         & $0.042 p$       \\
    Reference water pressure $p_{w,ref}$         & $0.042 p$       \\
    ASL thickness $l_b$, $\mu$m                  & 680             \\
    AAl thickness $l_a$, $\mu$m                  & 15              \\
    Cell active area, $S_{cell}$, cm$^2$         & 1.54            \\
    Parameter $\beta = b_h/b_w = \alpha_{w}/\alpha_h$ & {$1/3$~Ref.\cite{Zhu_06b}}  \\
    \hline
\end{tabular}
\end{center}
\caption{The working cell parameters reported by Shi \etal\cite{Shi_08,Shi_07}.
}
\label{tab:Shi_fixed}
\end{table}
\begin{table}
    \small
\begin{center}
\begin{tabular}{|l|c|}
\hline
    HOR Tafel slope $b_h$, mV / exp                               &  107  \\
    HOR transfer coefficient $\alpha_h = RT/(b_h F)$              &  0.864  \\
    DL volumetric capacitance $\Cdl$, F~cm$^{-3}$                 &  3.36  \\
    DL superficial capacitance $C_{dl,s}$, mF~cm$^{-2}$           &  $5.04$  \\
    HOR exchange current density $i_*$, A~cm$^{-3}$               &   $355^*$  \\
    AAL ionic conductivity $\sion$, mS~cm$^{-1}  $                &   $1.65$   \\
    ASL Knudsen H$_2$ diffusivity $D_{K,h}$, cm$^2$~s$^{-1}$      &   $0.807$   \\
    High--frequency resistance $R_{HFR}$, $\Omega$~cm$^2$         &   0.268  \\
    Ratio $\pwref/\phref$                                         &   0.723    \\
    ASL porosity/tortuosity                                       &   0.0761  \\
    Cable inductance $L_{cab}$, nH                                &   $418$ \\
\hline
\end{tabular}
\end{center}
\caption{
   The anode parameters resulted from fitting of the analytical model
   to the experimental spectrum in Figure~\ref{fig:Shi}.
   The parameter indicated by asterisk has been fixed.
}
\label{tab:Shi}
\end{table}
%




%

\subsection{Traps when using the Warburg finite-length element}

The Warburg finite-length element has been widely used in ECMs for fitting
the impedance spectra\cite{Leonide_09,Kromp_13,Huang_15} (see also a review\cite{Nielsen_14}
and the references therein).
The analytical model, Eq.\eqref{eq:tZanly} and Eq.\eqref{eq:cond} below enable
to indicate possible traps when using
this element. The condition for neglecting the pressure gradient
in the ASL is\cite{Kulikovsky_25a2}:
\begin{equation}
  \dfrac{J R T L}{F D_{K,h}\left(1 + \dfrac{p_c B_0(3 - 2 y_c)}{\mu D_{K,h}}\right)p_c} \ll 1
  \label{eq:cond}
\end{equation}
where $y_c$, $p_c$ are the hydrogen molar fraction and total pressure, respectively, in the channel.
Under this condition, the pressure growth toward the active layer produced
by the Knudsen diffusion and/or due to the finite hydraulic permeability of the porous
media can be neglected.

Care should be taken when using isobaric ECMs for fitting the spectra
measured at OCV. In cells with the YSZ electrolyte, the electronic leakage generates current on the order of
1--10 mA~cm$^{-2}$, depending on the electrolyte thickness\cite{Duncan_11}. Under certain
set of the ASL transport parameters, the condition \eqref{eq:cond} could be violated already
at OCV due to the leakage current. In this case, the effect of
the pressure gradient cannot be ignored and it is recommended using Eq.\eqref{eq:tZanly},
rather than the isobaric Warburg finite-length element.

\begin{figure}
    \begin{center}
        \includegraphics[scale=0.45]{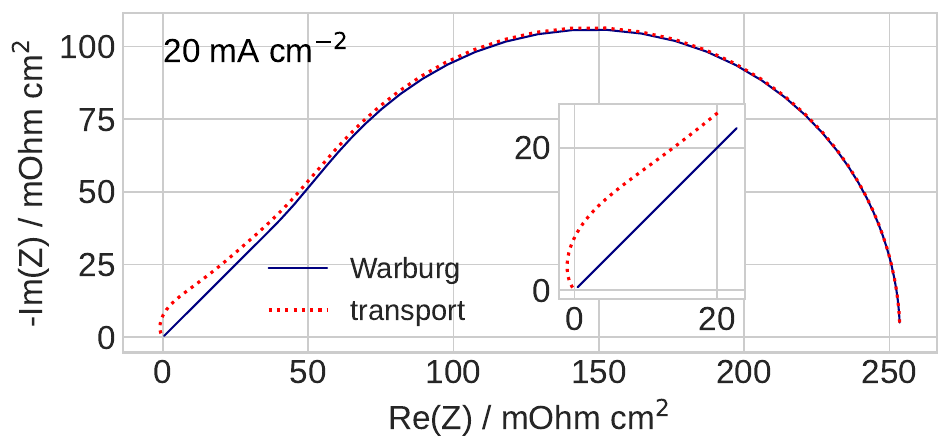} \\
        \caption{The Warburg finite--length impedance $Z_W/(qE_1)$,
        Eq.\eqref{eq:tZW}, (solid curve) and the transport impedance,
        Eq.\eqref{eq:tZtra}, (dotted curve)
        calculated with the parameters from Table~\ref{tab:parms}. The inset
        shows the zoomed high--frequency part.
        }
        \label{fig:ztrazw}
    \end{center}
\end{figure}

More specifically, at low to medium frequencies, the transport term
in Eq.\eqref{eq:tZanly} is close to the Warburg finite-length impedance (Figure~\ref{fig:ztrazw}).
Thus, the Warburg element can be safely used to determine the transport resistivity.
However, if Eq.\eqref{eq:cond}
does not hold, it is impossible to extract a correct ASL hydrogen diffusivity from the Warburg element.

Further, the active layer ionic conductivity $\sion$ manifests itself in the high--frequency
part of the impedance spectrum. The characteristic frequency $f_i$
of ionic transport in the active layer is\cite{Kulikovsky_20f}
\begin{equation}
    f_i = \dfrac{1.71 \sion}{\Cdl l_a^2}, \quad \text{Hz.}
    \label{eq:fi}
\end{equation}
With the parameters from Table~\ref{tab:parms}, $f_i \simeq 400$ Hz.
The high--frequency behavior of the Warburg impedance and
the transport term in Eq.\eqref{eq:tZanly} are very different
(Figure~\ref{fig:ztrazw}), and using the Warburg
element may return incorrect  $\sion$.

To summarize,
\begin{itemize}

\item The Warburg finite-length element can be safely used to
    derive the static transport resistivity.

\item The ASL effective hydrogen diffusivity can be correctly estimated from the fitted
    Warburg element only if Eq.\eqref{eq:cond} holds.

\item The Warburg element distorts the high--frequency part of the spectrum,
   which may result in an incorrect value of the fitted active layer ionic conductivity.

\item The validity of Eq.\eqref{eq:tZanly} is limited by Eq.\eqref{eq:cdion},
    which typically is much less restrictive than Eq.\eqref{eq:cond}.
    Using Eq.\eqref{eq:tZanly} for fitting OCV and low--current spectra is thus much safer
    than using ECMs with the Warburg element.

\end{itemize}

Finally, we note that Eq.\eqref{eq:cond} was derived for the H$_2$--H$_2$O mixture.
The presence of the third non-reacting component in the mixture (e.g. nitrogen)
would increase the pressure gradient in the ASL. The higher pressure gradient
would be required to achieve zero total flux of the non-reacting gas in the transport layer.
Thus, in ternary systems, the condition Eq.\eqref{eq:cond} for neglecting the
pressure gradient is necessary but not sufficient.

\section{Conclusions}

Analytical low-current and numerical high--current models for the anode impedance
of an anode--supported button SOFC operated with neat hydrogen
is developed. The models are based on the dusty gas transport model
and they take into account the diffusive and the pressure--gradient driven
transport of the hydrogen--water vapor mixture in the porous
anode support layer (ASL). Numerical tests show that for the standard
anode parameters, the analytical model, Eq.\eqref{eq:tZanly}, works well up to the cell current
density on the order of 20 mA~cm$^{-2}$.
The ASL transport resistivity is given by Eq.\eqref{eq:tRtra}.
We show that fitting the model which neglects the pressure gradient to an experimental spectrum
may strongly underestimate the effective hydrogen diffusivity of the anode.

Fitting of the developed analytical model
to the spectrum of a button--type SOFC measured at open circuit conditions returns
a set of reasonable anode transport and kinetic parameters.
The derived Eq.\eqref{eq:tZanly} for the anode impedance contains
the ASL Knudsen hydrogen diffusivity, hydraulic permeability, porosity/tortuosity
ratio, the active layer ionic conductivity, the DL capacitance
and the HOR Tafel slope. Theoretically,  all the aforementioned parameters
could be obtained from fitting the model to experimental spectra.
The analytical model reveals the traps when fitting the equivalent circuit models
with the Warburg finite--length element to the SOFC spectra.

\appendix

\section{Derivation of the analytical impedance, Eq.\eqref{eq:tZanly}}

\label{sec:deriv}

If the DC current is small, the factors $Y_1$, $P_1$ and $E_1$
can be approximated by constant values.
The terms $Y_1 y_1^1$, $P_1\tp_1^1$, $E_1$ in Eq.\eqref{eq:teta1x}
and  $\tN_1^1$ in Eq.\eqref{eq:hj1}
are thus independent of $\tx$. Here, $Y_1$, $P_1$ and $E_1$ are
calculated setting $\tp^0=\ph$, $y^0 = y_c$ in Eqs.\eqref{eq:YPE}.
Note that the perturbation amplitudes are still functions of $\tX$ (see below).

A good approximation of the overpotential $\teta^0$ can be obtained from the static
charge conservation equation, Eq.\eqref{eq:vccnum}, in the form
\begin{equation}
  \veps^2\pdr{\hj^0}{\tx} = - q\tp_c\,\bigl(y_c\exp(\teta^0)
             - r_{ref} (1 - y_c)\exp(- \beta\teta^0)\bigr)
   \label{eq:tvccj}
\end{equation}
where $\hj^0 = \pdra{\teta^0}{\tx}$ is the local ionic current density.
Assuming that the right side of Eq.\eqref{eq:tvccj} is constant and integrating
this equation over $\tx$ from 0 to 1, we get the current--voltage relation
\begin{equation}
   J = l_a i_* q \tp_c\,\bigl(y_c\exp(\teta^0) - r_{ref} (1-y_c)\exp(- \beta\teta^0)\bigr)
   \label{eq:vcc}
\end{equation}
Eq.\eqref{eq:vcc} is valid if the cell current density is small\cite{Kulikovsky_10c}:
\begin{equation}
   J \ll \dfrac{\sion b_h}{l_a}
   \label{eq:cdion}
\end{equation}
For the set of active layer parameters in Table~\ref{tab:parms}, the right side of Eq.\eqref{eq:cdion}
is 67 mA~cm$^{-2}$, i.e., under given cell temperature, the equations of this section are valid up to
the cell currents on the order of 10 mA~cm$^{-2}$.

With this assumption, Eq.\eqref{eq:teta1x} is an ODE with the constant coefficients
and the solution to Eq.\eqref{eq:teta1x} is
\begin{multline}
    \teta^1(\tx) = \dfrac{\hj^1\cosh(\phi \tx)}{\phi \sinh(\phi)}
        - \dfrac{q \left(Y_1 y_1^1 + P_1\tp_1^1\right)}{\ri \tom + q E_1}, \\
    \text{where}\quad \phi = \dfrac{1}{\veps}\sqrt{\ri \tom + q E_1}
    \label{eq:tetax_sol}
\end{multline}
The anode impedance is $\tZ_{anly} = \teta^1/\hj^1|_{\tx=1}$ and
since $\hj^1 = \tN_1^1 J_* / j_*$, from Eq.\eqref{eq:tetax_sol} we get
\begin{equation}
   \tZ_{anly} = \dfrac{1}{\phi \tanh(\phi)}
         - \dfrac{\chi \left(Y_1 y_1^1 + P_1\tp_1^1\right)}{\tN_1^1(\ri \tom + q E_1)},
   \label{eq:tZsa}
\end{equation}
where $\chi$ is given in Eq.\eqref{eq:phi}.

Further, at low DC currents, Eqs.\eqref{eq:tN1}, \eqref{eq:sumd1F}, \eqref{eq:hyd1F2} can be simplified.
Setting $\pdra{\tp^0}{\tX} = \pdra{y^0}{\tX} = 0$, $\tp^0 = \ph$, and $y^0=y_c$,
from Eqs.\eqref{eq:tN1}, \eqref{eq:sumd1F}, \eqref{eq:hyd1F2}
we get the reduced equations for the perturbation amplitudes $\tN^1$, $\tp^1$ and $y^1$:
\begin{align}
   & \psi^2\pdr{\tN^1}{\tX} = - \left(\ph y^1 + y_c\tp^1\right)\ri\tom,                \label{eq:tN1b}    \\
   &\left(1 + \ph (3 - 2 y_c)\right)\pdr{\tp^1}{\tX} = 2 \tN^1                         \label{eq:sumd1Fb} \\
   & 2\ph\pdr{y^1}{\tX} + \left(1 + 2y_c + 3\ph\right) \pdr{\tp^1}{\tX} = - 2 K \tN^1  \label{eq:hyd2Fb}
\end{align}

Eqs.\eqref{eq:tN1b}--\eqref{eq:hyd2Fb} form a system
of linear ODEs with constant coefficients. This system with the boundary conditions
Eq.\eqref{eq:bcs} can be solved analytically. The solution
is rather cumbersome and it is not displayed here. Setting $\tX=1$ in
the solutions we obtain the explicit formulas for $\tN_1^1$, $y_1^1$ and $\tp_1^1$,
which appear in the second (transport) term in  Eq.\eqref{eq:tZsa}.
After some algebra, this gives the formula for the analytical impedance $\tZ_{anly}$

\section{Transport coefficients and the dimensionless parameters}

\begin{table}
\begin{center}
\begin{tabular}{|l|}
\hline
  $j_* = \sion b_h / l_a$ \\
  $\tJ =  {J}/{J_*}, \quad J_* = {2 F \mu D_{K,h}^2}/{(R T L B_0)}$ \\
  $\hj  = j/j_*, \quad j_* = \sion b_h /l_a$ \\
  $K = D_{K,h} / D_m$ \\
  $\tN =  {N}/{N_*}, \quad N_* = {\mu D_{K,h}^2}/{(R T L B_0)}$ \\
  $\tp =  {p}/{p_*}, \quad p_* = {\mu D_{K,h}}/{B_0}$ \\
  $q =    {\mu D_{K,h}}/{(B_0 \phref)}$ \\
  $\tR_{tr} = R_{tr} l_a /\sion$ \\
  $r_{ref} = {\phref}/{\pwref}$ \\
  $\tit = {t}/{t_*}, \quad t_* = {\Cdl b_h}/{i_*}$ \\
  $\tX = {X}/{L}$ \\
  $\tx =  {x}/{l_a}$ \\
  $Y_1, P_1, E_1$, Eq.\eqref{eq:YPE} where $\tp^0 = \tp_1^0$, $y^0=y_1^0$, \\
  for numerical model, or $\tp^0 = \tp_c$, $y^0=y_c$  \\
  for analytical model \\
  $\tZ = Z \sion/l_a$ \\
\hline
  $\alpha = {R T}/{(b_h F)}$ \\
  $\beta   = b_h / b_w = {1}/{3}$  \\
  $\veps   = \sqrt{{\sion b_h}/{(i_* l_a^2)}}$ \\
  $\teta   = {\eta}/{b_h}$ \\
  $\teta^0$  Solution to Eq.\eqref{eq:vccnum} (numerical model), \\
  or solution to Eq.\eqref{eq:vcc} (analytical model) \\
  $\xi = (((3 - 2y_c) K + 3)\ph + K + 1)/(1 + \ph (3 - 2y_c))$ \\
  $\phi    = \sqrt{\ri\tom + q E_1}/ \veps$ \\
  $\chi    = {R T L \sion b_h}/{(2 F D_{K,h} \phref l_a)}$ \\
  $\psi    = \sqrt{\Cdl b_h D_{K,h}/ (i_* L^2)}$  \\
  $\tom = \omega \Cdl b_h / i_*$ \\
\hline
\end{tabular}
\end{center}
\caption{The dimensionless variables and parameters in alphabetic order.
}
\label{tab:dless}
\end{table}

The transport coefficients have been calculated as
\begin{equation}
   \begin{split}
      & B_0 = \dfrac{\lambda d^2}{32}, \quad \text{Ref.\cite{Bertei_15}} \\
      & D_{K,h} = \dfrac{\lambda d}{3} \sqrt{\dfrac{8 R T}{\pi M_{h}}} \\
      & D_m = \lambda D_m^{free}
   \end{split}
   \label{eq:tcoeffs}
\end{equation}
where
$\lambda$ is the porosity/tortuosity ratio,
$d$ the mean pore diameter (Table~\ref{tab:parms}).

The first derivative of Eq.\eqref{eq:y0x} over $\tX$ is given by
\begin{multline}
   \pdr{y}{\tX} = - \dfrac{W \tp_c y_c}{\left(\tp_c + W \tX\right)^2} \\
      - \dfrac{6 W \tp_c^2 + 2 \left(3 W^2 \tX + 2 K \tJ + W\right) \tp_c + 3 W^3\tX^2}
              {4 \left(\tp_c + W \tX\right)^2}
   \label{eq:dydx}
\end{multline}


\section*{Nomenclature}

\small

\begin{tabular}{ll}
    $\hat{}$     &  Marks the dimensionless local ionic current density \\
    $\tilde{}$   &  Marks the other dimensionless variables \\
    $B_0$        &  Hydraulic permeability, m$^2$      \\
    $b$          &  Tafel slope, V                      \\
    $d$          &  Mean pore diameter, m                             \\
    $\Cdl$       &  Double layer volumetric capacitance, F~m$^{-3}$      \\
    $\phref$     &  Reference hydrogen pressure, Pa      \\
    $\pwref$     &  Reference water vapor pressure, Pa      \\
    $D_{m}$      &  Effective binary H$_2$--H$_2$O molecular diffusion    \\
                 &  coefficient in the ASL, m$^2$~s$^{-1}$  \\
    $D_m^{free}$ &  Binary H$_2$--H$_2$O molecular diffusion    \\
                 &  coefficient in a free space, m$^2$~s$^{-1}$  \\
    $D_{K,h}$    &  Effective Knudsen diffusion coefficient \\
                 &  of hydrogen, m$^2$~s$^{-1}$  \\
    $E$          &  Dimensionless parameter, Eq.\eqref{eq:YPE}   \\
    $F$          &  Faraday constant, C~mol$^{-1}$                             \\
    $J$          &  DC cell current density, A~m$^{-2}$     \\
    $j$          &  Local ionic current density, A~m$^{-2}$     \\
    $K$          &  $ = D_{K,h} / D_m$         \\
    $L$          &  Anode support layer thickness, m     \\
    $l_a$        &  Anode active layer thickness, m     \\
    $M$          &  Molecular weight, kg~mol$^{-1}$ \\
    $N$          &  Molar flux of hydrogen, mol~m$^{-2}$~s$^{-1}$        \\
    $P$          &  Dimensionless parameter, Eq.\eqref{eq:YPE}   \\
    $p$          &  Pressure, Pa         \\
    $p_*$        &  Characteristic pressure, Pa, Eq.\eqref{eq:dless}    \\
    $q$          &  Dimensionless parameter, Eq.\eqref{eq:qrref}          \\
    $R$          &  Gas constant, J~K$^{-1}$~mol$^{-1}$  \\
    $r_{ref}$    &  Dimensionless parameter, Eq.\eqref{eq:qrref} \\
    $S_{cell}$   &  Cell active area, m$^2$  \\
    $T$          &  Cell temperature, K \\
    $W$          &  Dimensionless parameter, Eq.\eqref{eq:W}       \\
    $X$          &  Coordinate through the ASL   \\
                 &  counted from the chamber, m  \\
    $x$          &  Coordinate through the anode active layer  \\
                 &  counted from the ASL/AAL interface, m      \\
    $Y$          &  Dimensionless parameter, Eq.\eqref{eq:YPE}   \\
    $y$          &  Molar fraction of hydrogen    \\
    $Z$          &  Impedance, Ohm~m$^2$          \\ [1em]
\end{tabular}

\newpage

{\bf Subscripts:\\}

\begin{tabular}{ll}
    $*$       & Characteristic value \\
    $a$       & ASL/active layer interface \\
    $anly$    & Analytical (impedance) \\
    $c$       & Channel/ASL interface     \\
    $h$       & Hydrogen \\
    $K$       & Knudsen diffusion \\
    $RC$      & Parallel $RC$--circuit impedance \\
    $tr$      & Transport \\
    $m$       & Molecular diffusion \\
    $W$       & Warburg impedance   \\
    $w$       & Water             \\[1em]
\end{tabular}

{\bf Superscripts:\\}

\begin{tabular}{ll}
    $0$      & Steady--state value \\
    $1$      & Small--amplitude perturbation \\[1em]
\end{tabular}

{\bf Greek:\\}

\begin{tabular}{ll}
    $\beta$         &  Dimensionless parameter, Eq.\eqref{eq:betaveps} \\
    $\veps$         &  Dimensionless parameter, Eq.\eqref{eq:betaveps} \\
    $\eta$          &  HOR overpotential, positive by convention, V \\
    $\lambda$       &  ASL Porosity/tortuosity ratio  \\
    $\mu$           &  Mixture dynamic viscosity, Pa~s   \\
    $\sion$         &  AAL ionic conductivity, $\Omega^{-1}$~m$^{-1}$ \\
    $\phi$          &  Dimensionless parameter, Eq.\eqref{eq:phi} \\
    $\chi$          &  Dimensionless parameter, Eq.\eqref{eq:phi} \\
    $\psi$          &  Dimensionless parameter, Eq.\eqref{eq:psi} \\
    $\omega$        &  Angular frequency, s$^{-1}$                \\
\end{tabular}

\newpage

\end{document}